\documentclass[12pt]{iopart}
\usepackage{graphicx}
\usepackage{color}
\usepackage{iopams}
\usepackage{setstack}
\begin{document}
\title{Matter Power Spectrum for Convex Quartessence}
\author{R. R. R. Reis\dag\ , M. Makler\dag \ddag\ and I. Waga\dag}
\address{\dag\ Universidade Federal do Rio de Janeiro, Instituto de F\'\i
sica, CEP 21941-972, Rio de Janeiro, RJ, Brazil}
\address{\ddag\ Observat\'orio Nacional -- MCT, CEP 20921-400, Rio de
Janeiro, RJ, Brazil}

\begin{abstract}

The possibility of unifying dark-matter and dark-energy has
recently attracted considerable interest. In this so called
quartessence scenario, a single component is responsible for both
the clustering of matter and the accelerated expansion of the
universe. A model archetype for such scenario is provided by the
Chaplygin gas. Although this model is in agreement with the data
on the expansion history, problems arise in the power spectrum of
density fluctuations for adiabatic perturbations. In this
contribution we consider other quartessence models and confirm
that instabilities and oscillations in the matter power spectrum
are a characteristic of more generic quartessence models, namely
those with a convex equation of state. We show that, as in the
Chaplygin case, this kind of problem can be solved by considering
intrinsic non-adiabatic perturbations such that, as an initial
condition, the perturbed fluid is gradient pressure free. We also
discuss how the problems of adiabatic quartessence can be
circumvented by other types of equations of state.

\end{abstract}

\maketitle

\section{Introduction}

The twentieth century witnessed the establishment of a
cosmological model consistent with a number of astronomical
observations. In this \textit{standard} model, the dynamics of the
universe is governed by the local distribution of matter-energy,
following Einstein's theory of gravitation. On large scales the
universe is well described within a nearly flat
Friedmann--Lema\^{\i}tre--Robertson--Walker (FLRW) cosmological
scenario, and is presently undergoing an accelerated expansion.
Despite the success of this model, several key features of the
universe remain unclear. For example, what constitutes the
dark-matter (DM) that would trigger gravitational clustering, and
the dark-energy (DE) that is supposed to power the accelerated
expansion? Are these two dark components separate entities or
different manifestations of a single matter-content?

Although there are some candidates for the DM from particle
physics \cite{DM}, there is yet no evidence for these particles in
laboratory experiments. For the DE the situation is even more
puzzling since there is no natural candidate from fundamental
physics, although the cosmological constant and a dynamical scalar
field are the most popular \cite{denergy}. Thus, the current
information about DE and DM comes solely from astronomical
observations. From a simplicity viewpoint it is interesting to
investigate the possibility of describing the phenomenology
associated with both DE and DM through a single matter-component,
\textit{unifying dark matter} (UDM) --- or simply quartessence ---
\cite{makler03}, reducing to one the unknown components of the
material substratum of the universe.

A few candidates for such UDM appeared in the literature in recent
years, the most popular being the generalized Chaplygin gas
\cite{kamenshchik},\cite{tesemartin},\cite{bilic,fabris01,BentoGCG},
in what we call the Quartessence Chaplygin Model (QCM). For this
specific model, several confrontations with observational data
were performed and constraints were set on the model parameters.
The generalized Chaplygin gas (as quartessence) appears to be
compatible with all available data regarding the expansion history
(see e.g. ref. \cite{makler03b}). However, for adiabatic
perturbations, it was shown that the mass power spectrum presents
strong oscillations and instabilities, allowing only models very
close to the $\Lambda$CDM limit \cite{sandvik02}. This problem can
be avoided if non-adiabatic perturbations with a specific initial
condition ($\delta p=0$) are considered \cite{reis03b}. In
\cite{sandvik02} it was argued that these oscillations and
instabilities would be present in any quartessence model such that
$p=p(\rho)$. We argue that this is indeed the case for convex
equations of state. This is illustrated through the investigation
of two extreme models of convex quartessence, one with a very
steep and another with a very gentle equation of state. We show
that for these models, as in the QCM case, the above mentioned
problems in the mass power spectrum can always be avoided with the
same kind of entropy perturbation. On the other hand, we argue
that the problems of adiabatic quartessence could be avoided by
other types of equations of state.

The paper is organized as follows. In section \ref{pheno4ess} we
review the QCM and introduce the two mentioned extreme
quartessence models. The treatment of linear perturbations in
these models is discussed in section \ref{pert}. The computation
of the power spectrum in the adiabatic and nonadiabatic cases and
comparison with observational results are developed in section
\ref{pspec}. Finally, we sum up our results and present concluding
remarks in section \ref{disc}.

\section{Phenomenological Models of Quartessence\label{pheno4ess}}

In this work we model the UDM component as a fluid with isotropic
energy-momentum tensor, whose only independent components are the
pressure $p$ and the density $\rho$. We further assume that the
space averaged dynamical variables are related by the equation of
state (EOS) $\bar{p}=\bar{p}\left( \bar{\rho}\right)  $, i.e. the
background is a perfect fluid. For this EOS to be a quartessence
candidate, it has to fulfill some conditions. For example,
neglecting the contribution of baryons, in the FLRW model, for the
universe to undergo acceleration, one must have at present $\left(
\bar{\rho}+3\bar{p}\right)  <0$. Thus we require $\bar{p}$ to be
sufficiently negative when $\bar{\rho}$ is small. On the other
hand, in the past, for the UDM component to behave like dark
matter we must have $\bar {\rho}\gg\bar{p}$, such that the
universe was decelerating and the density scaled as $a^{-3}$
(where $a$ is the scale factor). A remarkable property of
quartessence is that the weak energy condition is never violated
in such models. From the energy conservation equation
\begin{equation}
\overset{\cdot}{\bar{\rho}}=-\left(  \bar{\rho}+\bar{p}\right)
3\frac{\dot {a}}{a}=-\bar{\rho}\left(  1+w\left( \bar{\rho}\right)
\right)  3\frac {\dot{a}}{a}\;. \label{rhop}
\end{equation}
it follows that the line $\bar{p}=-\bar{\rho}$ cannot be crossed.
Since at early times we should have $w\approx0$, a straightforward
prediction of these models is, therefore, that $w>-1$. If
observations find that $w<-1$, this automatically rules out
quartessence models in which the background is a prefect fluid
(i.e. in which $w$ is a function only of $\bar{\rho}$) and where
the dark sector is not coupled to other components.

A simple example of an equation of state satisfying the above
conditions is given by an inverse power law
\cite{kamenshchik},\cite{tesemartin},\cite{bilic,fabris01,BentoGCG},
\begin{equation}
\bar{p}=-\frac{M^{4\left(  \alpha+1\right) }}{\bar{\rho}^{\alpha}}
\,\,\;\;\;\; \hbox{(Chaplygin Quartessence)}, \label{eostoy}
\end{equation}
where $M$ has dimension of mass and $\alpha$ is a dimensionless
parameter. The energy conservation equation (\ref{rhop}) has a
simple analytic solution for this EOS
\begin{equation}
\bar\rho_{\mathrm{Ch}}=\bar\rho_{\mathrm{Ch}0}\left[  (1-A)\left(
\frac
{a_{0}}{a}\right)  ^{3(\alpha+1)}+A\right]  ^{1/(\alpha+1)}. \label{rhoa}%
\end{equation}
Here $a_{0}$ is the present value of the scale factor and
$A=(M^{4}/\bar \rho_{\mathrm{Ch}0})^{(\alpha+1)}$. As expected,
when $a/a_{0}\ll1$, we have $\bar\rho_{\mathrm{Ch}}\propto a^{-3}$
and the fluid behaves as CDM. For late times, $a/a_{0}\gg1$, and
we get $\bar p_{\mathrm{Ch}}=-\bar\rho_{\mathrm{Ch}
}=-M^{4}=const.$ as in the cosmological constant case.

The QCM has been extensively discussed in the literature recently,
both setting observational limits to the model and studying its
properties and physical motivation. Nevertheless, since the QCM is
a prototypical and particularly simple example of quartessence we
shall briefly review a few results concerning the parameters of
the model. This will help to fix the notation and gain some
intuition, as several aspects are similar in generic models of
quartessence.

Since in QCM the equation of state parameter $w$ is a monotonic
decreasing function, it is easy to see from equation (\ref{rhop})
that the equation of state goes asymptotically to a cosmological
constant form $\bar{p}\left( \bar{\rho}\right)  =-\bar{\rho}$. The
minimum value of the density, obtained from the solution of this
equality, will be denoted by $\bar{\rho}_{\min}$ and is simply
given by $M^{4}$ in the QCM. Naturally, the current density
$\bar{\rho}_{\mathrm{Ch}0}$ cannot be lower than $\rho_{\min}$
(otherwise, as the $\bar{p}=-\bar{\rho}$ line cannot be crossed,
QCM would always have had $w<-1$, never acting as dark-matter).
Thus $\bar{\rho}_{min}<\bar{\rho }_{\mathrm{Ch}0}<\infty$, which
restricts the parameter $A$ to the range $0<a<1$. Clearly for
$A=0$ the fluid behaves as dust, while for $A=1$ it acts as a
cosmological constant.

Now, let us discuss the range of allowed values for $\alpha$. In
most of the analyses in the literature it is assumed that
$0\leq\alpha\leq1$. These limits are imposed because adiabatic
perturbations of the QCM are implicitly assumed. Notice that, for
these models, there is an absolute maximum value for the adiabatic
sound speed $c_{s}^{2}=\left(  \partial\bar{p}/\partial\bar{\rho
}\right)  $, which occurs for $\bar{\rho}=\bar{\rho}_{\min}$ and
is given by $c_{s\max}^{2}=\alpha$. Therefore, to avoid
superluminal propagation of signals the upper limit
($\alpha\leq1$) is imposed in the adiabatic case. Also, demanding
stability of adiabatic perturbations implies $\left(
\partial\bar{p}/\partial\bar{\rho}\right)  \geq0$, which requires $\alpha
\geq0$. However, as shown by Sandvik \textit{et al.}
\cite{sandvik02}, adiabatic QCM is ruled out by current data
unless $\alpha$ is very close to zero (corresponding to the
$\Lambda$CDM limit). Nevertheless, if entropy perturbations such
that the effective sound speed vanishes are allowed, the model is
again consistent with observational data \cite{reis03b}. In this
nonadiabatic case, the above limits on $\alpha$ are unnecessary.
For instance, if $\alpha>1$ there is no problem of superluminal
propagation for these specific entropy perturbations and even
models with $\alpha<0$ can, in principle, be in accordance with
observations. Since for quartessence we should have $w\approx0$ as
$\bar{\rho}\rightarrow\infty$, values of $\alpha<-1$ are not
allowed. Therefore, here we only require that $\alpha>-1$.

The QCM power spectrum exhibits strong oscillations or
instabilities, when adiabatic perturbations are considered and
$\alpha$ is not very close to zero. As we shall see in the next
section, this is due to the development in the fluid of a non null
adiabatic sound speed at recent times. For QCM the adiabatic sound
speed is related to the
equation of state parameter by $c_{s}^{2}=\alpha(M^{4}/\bar{\rho}%
)^{(\alpha+1)}=-\alpha w$. To avoid such oscillations in linear
scales, the parameter $\alpha$ of the QCM has to be restricted to
$\left\vert \alpha\right\vert <10^{-5}$, very close to the
$\Lambda$CDM limit \cite{sandvik02}. One may wonder if a steeper
variation of the equation of state would reduce the oscillations,
allowing a wider range of parameters, distinct from the
$\Lambda$CDM model. Let us consider for example an exponential
equation of state
\begin{equation}
\bar{p}=-M^{4}\exp\left(  -\frac{\alpha\bar{\rho}}{M^{4}}\right)
\,\,\;\;\;\;\hbox{(Exponential Quartessence)}. \label{exp4ess}
\end{equation}
In this case, the adiabatic sound speed has an additional
exponential suppression, $c_{s}^{2}=\alpha\exp\left(
-\alpha\bar{\rho }/M^{4}\right)  $. Thus, one could expect that
high ratios of $\bar{\rho }/M^{4}$ would attenuate the
oscillations in the power spectrum, without imposing strong
constraints on the parameter $\alpha$. However, the equation of
state parameter $w$ also has the same additional exponential
factor and, if the ratio $\bar{\rho}/M^{4}$ is always high, the
Universe does not enter in an accelerated phase. Therefore, as it
will be shown explicitly in the next section, this modification in
the equation of state does not help in solving the problem and
indicates that it is a general property of the model.

Before introducing our third type of quartessence, let us discuss
some noteworthy values of the exponential model parameters. First,
for Eq. (\ref{exp4ess}) to be a quartessence candidate
($w\approx0$ as $\bar{\rho }\rightarrow\infty$) the parameter
$\alpha$ must be non negative. The condition $\alpha\geq0$ also
guarantees stability for adiabatic fluctuations.
The minimum energy density of this model is given by $\bar{\rho}_{\min}%
=M^{4}W\left(  \alpha\right)  /\alpha$, where $W\left(  x\right)
$ is the Lambert function defined by $W\left(  x\right)
\exp\left(  W\left(  x\right) \right)  =x$ \cite{cranmer}.
Therefore, the maximum adiabatic sound speed is
$c_{s\max}^{2}=W\left(  \alpha\right)  $.

Another extreme case, such that the model has a $\Lambda$CDM
limit, is given by:
\begin{equation}
\bar{p}=-\frac{M^{4}}{\left[  \ln\left(  \bar{\rho}/M^{4}\right)
\right]
^{\alpha}}\,\,\,\;\;\;\;\hbox{(Logarithmic Quartessence)}. \label{log4ess}%
\end{equation}
In this case, for $w<0$ to be an increasing function of
$\bar{\rho}$, either $\alpha\geq0$ or $\alpha<-e$. The minimum
density is given by $\bar{\rho }_{\min}=M^{4}\exp\left(  \alpha
W\left(  \alpha^{-1}\right)  \right)  $ and
$c_{s\max}^{2}=1/W\left(  \alpha^{-1}\right)  $. As in the
exponential and QCM cases, for $\alpha=0$ we recover the $\Lambda
$CDM model for the background.

In the adiabatic case, where the condition $\alpha>0$ is fulfilled
such that $c_{s}^{2}>0$, the three models discussed above have a
common property, namely that the equation of state is convex, i.e.
$d^{2}p/d\rho^{2}=dc_s^2/d\rho<0$. Therefore, the maximum value of
the sound speed is reached at the minimum value of the density. We
shall refer to models that share this property as convex
quartessence. The above new \textit{Anz\"{a}tze} (Eqs.
\ref{exp4ess} and \ref{log4ess}) cover two extreme cases of convex
quartessence equations of state, one with a very steep and other
with a very gentle variation of the pressure with the density. We
believe that by analyzing these two examples we may have
indications of generic behaviors for any (monotonic, well behaved,
smooth, etc.) convex quartessence EOS. In the next sections we
shall focus on the matter power spectrum of these models, show
that they have the same oscillations as the Chaplygin case and
check that the introduction of nonadiabatic fluctuations solves
this problem.

\section{Evolution of Linear Perturbations\label{pert}}

The relativistic equations that govern the linear evolution of
scalar perturbations in a multi-component fluid, in the
synchronous gauge, are \cite{kodama, malik}
\begin{eqnarray}
\displaystyle\delta_{i}^{\prime}+3(c_{si}^{2}-w_{i})\frac{a^{\prime}}{a}%
\delta_{i}  &  =-(1+w_{i})\left(
kv_{i}+\frac{h_{L}^{\prime}}{2}\right)
-3w_{i}\frac{a^{\prime}}{a}\Gamma_{i},\label{eq1}\\
v_{i}^{\prime}+(1-3c_{si}^{2})\frac{a^{\prime}}{a}v_{i}  &  =\frac{c_{si}^{2}%
}{1+w_{i}}k\delta_{i}+\frac{w_{i}}{1+w_{i}}k\Gamma_{i},\label{eq2}\\
h_{L}^{\prime\prime}+\frac{a^{\prime}}{a}h_{L}^{\prime}  &  =-\sum
_{i}(1+3c_{si}^{2})8\pi G\bar{\rho}_{i}a^{2}\delta_{i}-24\pi
Ga^{2}\sum
_{i}\bar{p}_{i}\Gamma_{i}, \label{eq3}%
\end{eqnarray}
where $\delta_{i}$, $v_{i}$, and $\Gamma_{i}$ are, respectively,
the density contrast, the velocity perturbation, and the entropy
perturbation of component $i$, $h_{L}$ is the trace of the metric
perturbation, and $k$ is the commoving wave number. For the sake
of simplicity, we assumed that both the spatial curvature and the
anisotropic pressure perturbation vanish and that the
energy-momentum tensor of each component is separately conserved.
In the equations above the prime denotes derivative with respect
to conformal time
($dt=a\;d\eta$) and, as usual, $c_{si}^{2}=\bar{p}_{i}^{\prime}/\bar{\rho}%
_{i}^{\prime}$ and $w_{i}=\bar{p}_{i}/\bar{\rho}_{i}$ are,
respectively, the adiabatic sound speed and the equation of state
parameter of component $i$.

We will consider only two components in the universe: baryons ($\bar{p}_{b}%
=0$) and a quartessence component. To glimpse general features of
quartessence, we consider the two extreme models introduced above:
exponential (Eq. \ref{exp4ess}) and the logarithmic quartessence
(Eq. \ref{log4ess}). As remarked, these models have the
$\Lambda$CDM model as a limiting case for the background when
$\alpha=0$. At first and higher orders in perturbation theory,
however, these quartessence models with $\alpha=0$, have distinct
behavior as compared to that of dark-matter in $\Lambda$CDM
\cite{reis04,FabrisLCDM}.

The independent conservation of energy-momentum for baryons and
quartessence leads to
\begin{equation}
\bar{\rho}_{b}=\bar{\rho}_{b0}a^{-3}. \label{eq10}%
\end{equation}
For the exponential and logarithmic cases, the energy conservation
equation does not have an analytical solution. We rewrite Eq.
(\ref{rhop}) as
\begin{equation}
\frac{du}{da}+3\frac{u}{a}\left(  1+w\right)  =0, \label{econsu}%
\end{equation}
were $u=\bar{\rho}/\bar{\rho}_{0}$,
$w=w\left(u,\alpha,\tilde{A}\right) $, and
$\tilde{A}=M^{4}/\bar{\rho}_{0}$. For the exponential case one has
\begin{equation}
w=-\frac{\tilde{A}}{u}\exp\left(  -\frac{\alpha
u}{\tilde{A}}\right)
\end{equation}
whereas the logarithmic case leads to
\begin{equation}
w=-\frac{\tilde{A}}{u}\left(  \ln\frac{u}{\tilde{A}}\right)
^{-\alpha}.
\end{equation}
Therefore, given $\alpha$ and $\tilde{A}$ we numerically solve
(\ref{econsu}) with the
initial condition $u(1)=1$ to get $u(a)$ and then $\bar{\rho}=\bar{\rho_{0}}%
u$. Notice that, for each $\alpha$, $\tilde{A}$ has to be chosen
such that the condition $0\leq\bar{\rho}_{\min}/\rho_{0}\leq1$ is
fulfilled.

Let us write Eqs. (\ref{eq1}-\ref{eq3}) for the
baryon-quartessence system. For baryons, as $c_{sb}^{2}=0$ and
$w_{b}=0$, Eq. (\ref{eq2}) implies $v_{b}\propto a^{-1}$. As a
matter of simplification, we take $v_{b}=0$. Now Eq (\ref{eq1})
implies $h_{L}^{\prime}/2=-\delta_{b}^{\prime}$. Using this result
in Eq. (\ref{eq3}) and changing the \textquotedblleft
time\textquotedblright\ variable, we obtain, for the flat case
($\Omega
_{b0}+\Omega_{q0}=1$),%
\begin{equation}
\frac{d^{2}\delta_{b}}{da^{2}}+\left(  \frac{2}{a}+\frac{\ddot{a}}{\dot{a}%
^{2}}\right)
\frac{d\delta_{b}}{da}=\frac{3H_{0}^{2}}{2\dot{a}^{2}}\left\{
\Omega_{b0}a^{-3}\delta_{b}+(1-\Omega_{b0})u\left[  \left(  1+3c_{sq}%
^{2}\right)  \delta_{q}+3w_{q}\Gamma_{q}\right]  \right\}  . \label{syst1}%
\end{equation}
For the quartessence component, Eq. (\ref{eq1}) can be rewritten
as
\begin{equation}
\frac{d\delta_{q}}{da}+3\left(  c_{sq}^{2}-w_{q}\right)  \frac{\delta_{q}}%
{a}=-(1+w_{q})\left(
\frac{kv_{q}}{\dot{a}a}-\frac{d\delta_{b}}{da}\right)
-3\frac{w_{q}}{a}\Gamma_{q}, \label{syst2}%
\end{equation}
whereas Eq. (\ref{eq2}) gives%
\begin{equation}
\frac{dv_{q}}{da}+(1-3c_{sq}^{2})\frac{v_{q}}{a}=\frac{k}{1+w_{q}}\frac
{1}{a\dot{a}} \left(c_{sq}^{2}\delta_{q}+w_{q}\Gamma_{q}\right)  , \label{syst3}%
\end{equation}
The derivatives of the scale factor with respect to $t$ are given
by
\begin{eqnarray}
\dot{a}  &  =H_{0}\sqrt{\frac{\Omega_{b0}}{a}+(1-\Omega_{b0})\;u\;a^{2}%
},\label{eq25}\\
\ddot{a}  &  =-\frac{H_{0}^{2}}{2}\left(  \frac{\Omega_{b0}}{a^{2}}%
+(1-\Omega_{b0})\;u\;(1+3w_{q})\;a\right)  . \label{eq26}%
\end{eqnarray}
In the next section we apply these results to compute the matter
power spectrum for quartessence models.

\section{The Power Spectrum}

\label{pspec}

In order to obtain power spectra for the considered models we
evolved Eqs.\ (\ref{syst1}-\ref{syst3}) from $z=500$ to $z=0$,
using the solutions of Eq. (\ref{econsu}). At $z=500$ the
quartessence fluid behaves like CDM and, to take this into
account, proper initial conditions were established: we assumed
$v_{q}=0$, a scale invariant primordial spectrum, and used the
BBKS transfer function \cite{bardeen86}, with the following
effective shape parameter \cite{sugiyama95,beca},
\begin{equation}
\Gamma_{eff}=(\Omega_{b0}+\Omega_{m}^{eff})h\;\exp\left(  -\Omega_{b0}%
-\frac{\sqrt{2h}\;\Omega_{b0}}{\Omega_{b0}+\Omega_{m}^{eff}}\right)
,
\end{equation}
where the term
$\Omega_{m}^{eff}=(1-\Omega_{b0})\lim_{a\rightarrow0}(ua^{3})$ is
the effective matter density.

\begin{figure}[ptb]
\includegraphics[height= 6.5cm,width=8cm]{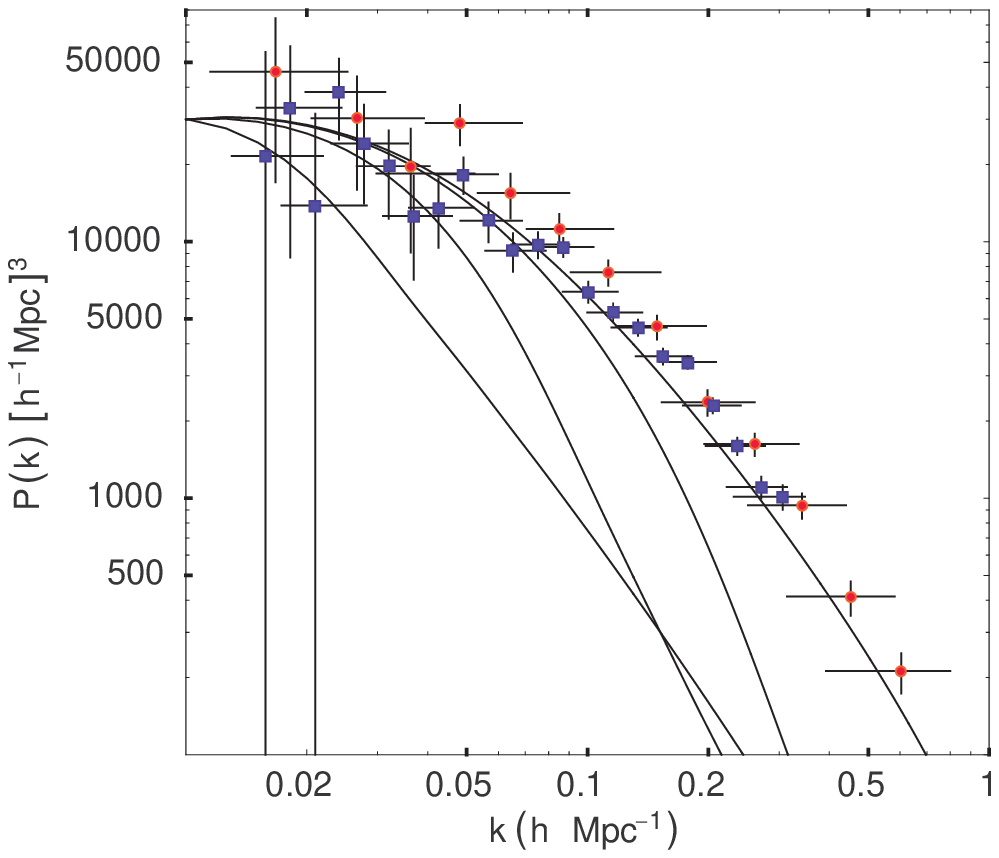}
\includegraphics[height= 6.5cm,width=8cm]{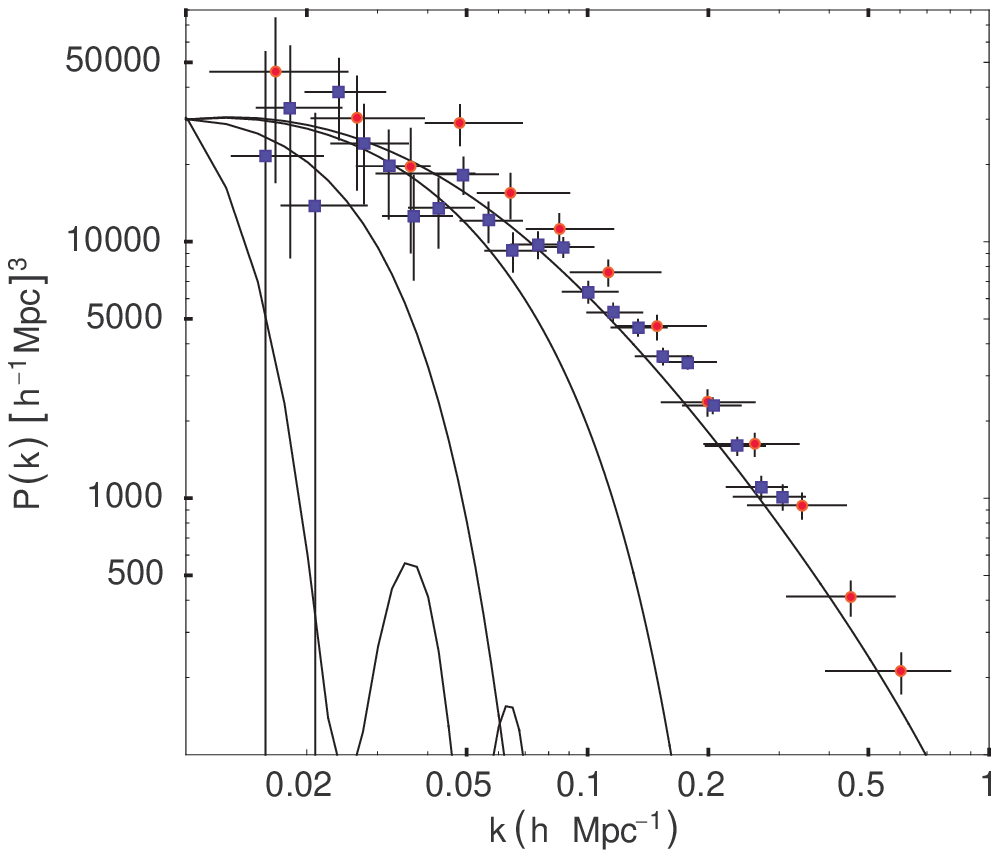}\caption{Baryon (left
panel) and total (right panel) mass power spectra for exponential
quartessence, in the adiabatic case. The curves from top to bottom
correspond to $\alpha=0,10^{-5},10^{-4},10^{-3}$ and $M^{4}/
\bar{\rho}_{0}$ is chosen such that all models have
$\Gamma_{eff}=0.18 $. The (blue) squares are the data points from
the Sloan Digital Sky Survey as compiled in \cite{SDSSWMAP} and
the (red)
circles are from the 2dF galaxy redshift survey as compiled in \cite{2dF}.}%
\label{expPk}
\end{figure}

\begin{figure}[ptb]
\includegraphics[height= 6.5cm,width=8cm]{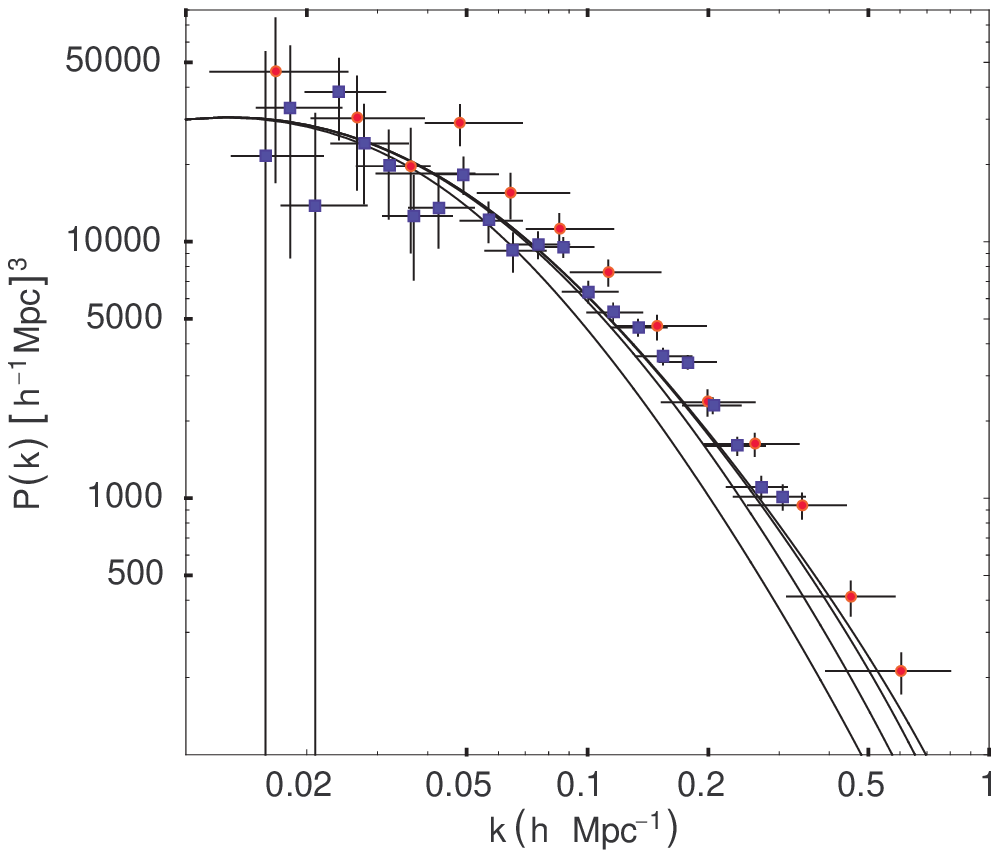}
\includegraphics[height= 6.5cm,width=8cm]{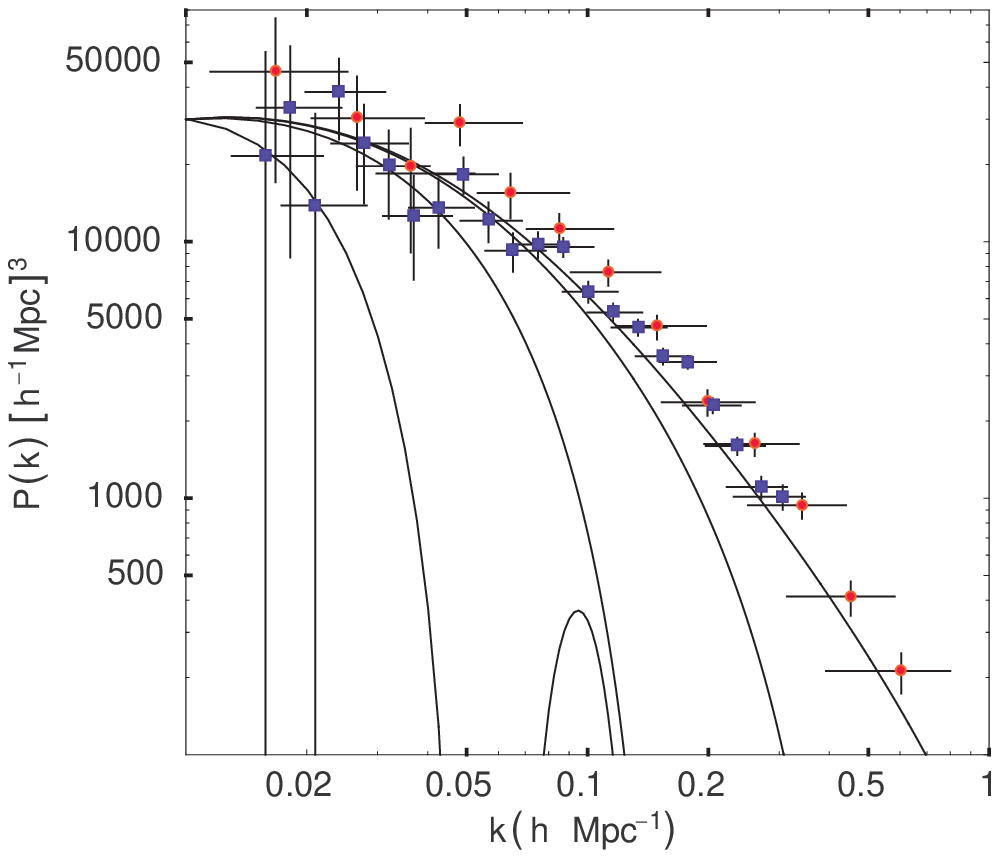}\caption{Baryon (left
panel) and total (right panel) mass power spectra for logarithmic
quartessence, in the adiabatic case. The curves from top to bottom
correspond to $\alpha=0,10^{-5},10^{-4},10^{-3}$ and $M^{4}/
\bar{\rho}_{0}$ is chosen such that all models have
$\Gamma_{eff}=0.18 $. The (blue) squares are the data points from
the Sloan Digital Sky Survey as compiled in \cite{SDSSWMAP} and
the (red)
circles are from the 2dF galaxy redshift survey as compiled in \cite{2dF}.}%
\label{logPk}
\end{figure}

In Fig. \ref{expPk} we present the baryon and the total mass power
spectra, for the exponential quartessence model, in the adiabatic
case ($\Gamma_{q}=0$), for $\alpha=0,10^{-5},10^{-4},10^{-3}$
(from top to bottom), $h=0.7$, $\Omega_{b0}=0.04$, and we choose
$M^{4}/\bar{\rho}_{0}$ such that all models have
$\Gamma_{eff}=0.18$. The total power spectrum can be written as
$P(k)\propto|\Omega_{b}\delta_{b}+\Omega_{q}\delta_{q}|^{2}$. The
normalization of the mass power spectra is arbitrarily fixed at
$k=0.01$ h Mpc$^{-1}$. The (blue) squares are data points from the
Sloan Digital Sky Survey (SDSS) as compiled in \cite{SDSSWMAP},
whereas the (red) circles correspond to the Two Degree Field
Galaxy Redshift Survey (2dFGRS) from \cite{2dF}. The presence of
oscillations in the total power spectra, due to a non-vanishing
adiabatic sound speed of the quartessence component, is apparent
in the figure. Further, even for the baryonic component, the
spectrum is compatible with the data only for $\alpha\ll 10^{-5}$.
In Fig. \ref{logPk}, we display the baryon and total mass power
spectra for logarithmic quartessence, for the same parameter
values as in Fig. \ref{expPk}. Again, oscillations and power
suppression are present in the total power spectrum. However, for
$\alpha \lesssim 10^{-4}$, the baryonic power spectrum is still
compatible with the data.

Notice that the amplitude of the initial power spectrum could be
reescaled such that the baryon spectrum provides a better fit to
the galaxy data. However, the normalization of the total power
spectrum is constrained by
gravitational lensing observations. Therefore, as remarked in \cite{sandvik02}%
, it is not possible to reconcile the constraints from both the
galaxy power spectrum and gravitational lensing observations for
non negligible values of $\alpha$ and adiabatic perturbations.

As discussed in \cite{reis03b} the oscillations (for
$c_{s}^{2}>0$) and instabilities (for $c_{s}^{2}<0$) in the
quartessence mass power spectrum that are present in the adiabatic
case, have their origin in a non-vanishing value of the right-hand
side in Eq. (\ref{eq2}). To avoid these effects we choose, for the
quartessence component
\begin{equation}
\Gamma_{q}=-\frac{c_{sq}^{2}}{w_{q}}\delta_{q}=-c_{sq}^{2}\frac{\delta\rho
_{q}}{p_{q}}. \label{condition}%
\end{equation}
>From the definition of entropy perturbation \cite{kodama}, we have
\begin{equation}
\delta p=p\Gamma+c_{s}^{2}\delta\rho. \label{def}%
\end{equation}
Thus the condition (\ref{condition}) is equivalent to set $\delta
p_{q}=0$. Moreover, it is sufficient that relation
(\ref{condition}) be satisfied at some initial time $t$, since
this condition is preserved along the linear evolution. This can
be easily seen by differentiating (\ref{def}) with respect
to time and substituting (\ref{condition}), which implies that $d\delta{p_{q}%
}/dt=0$, with no consideration about the exact form of the fluid
equation of state.

\begin{figure}[ptb]
\includegraphics[height= 6.5cm,width=8cm]{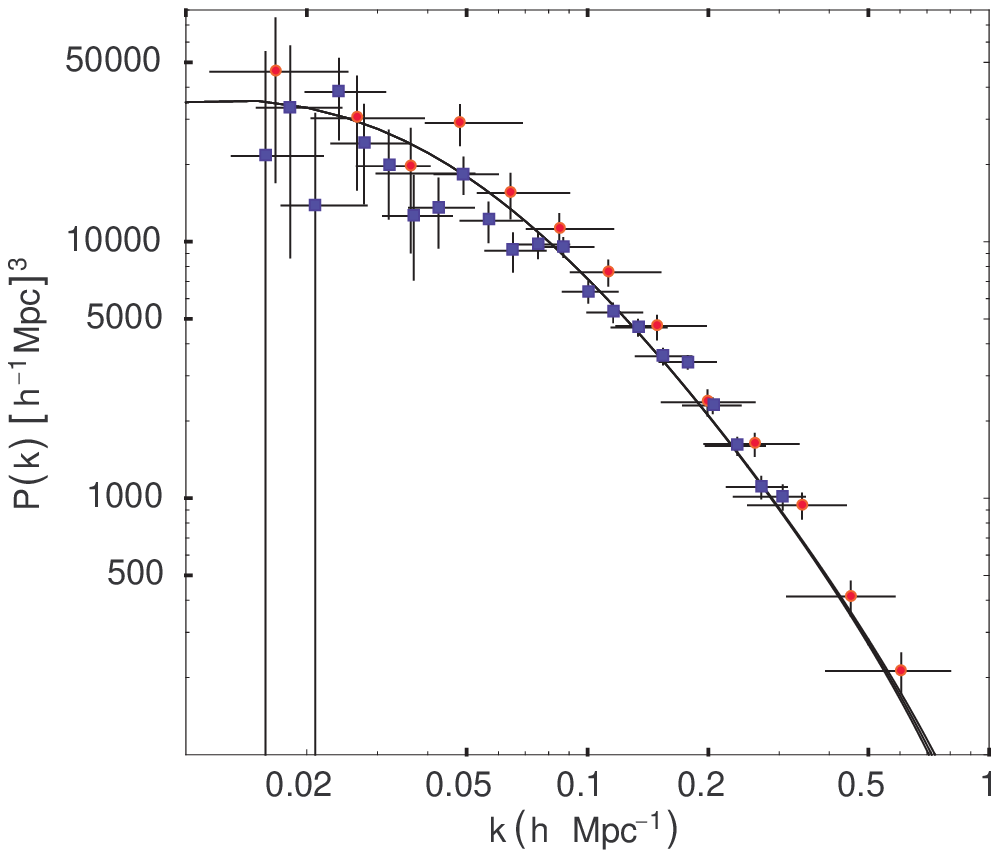}
\includegraphics[height= 6.5cm,width=8cm]{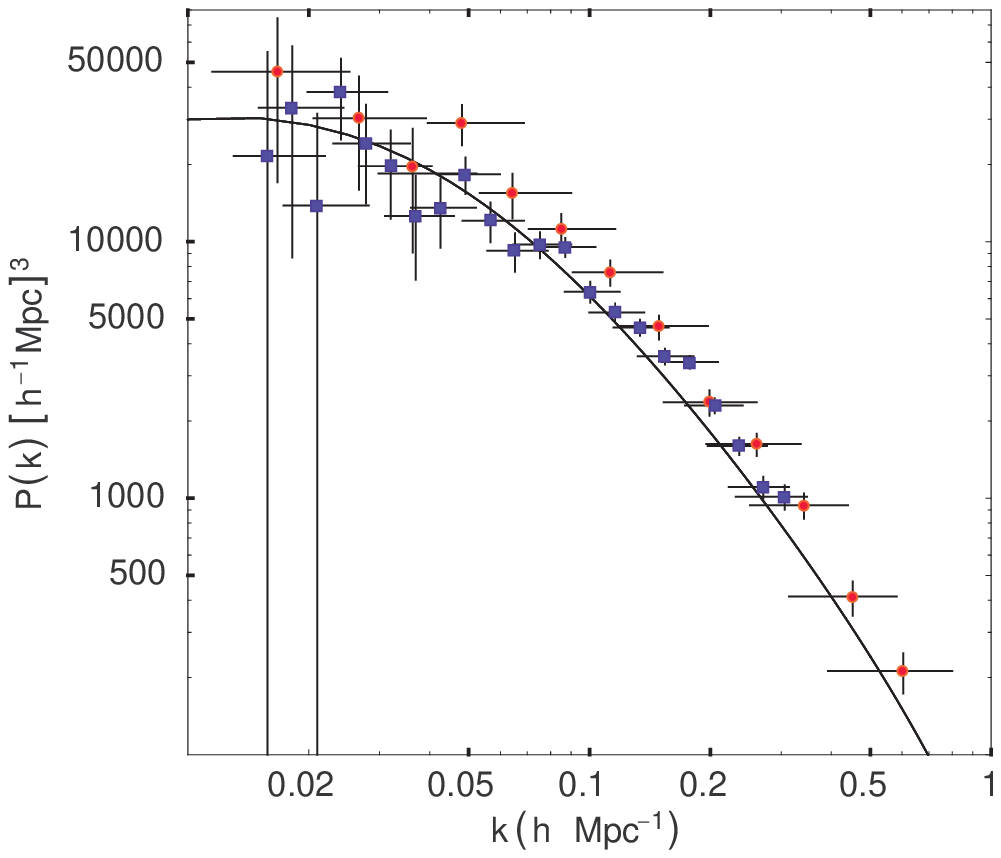}\caption{Baryon
mass power spectra for the model with exponential quartessence
(left panel) and logarithmic quartessence (right panel), in the
non-adiabatic case, $\alpha=0,0.1,0.2$ and $M^{4}/\bar{\rho}_{0}$
chosen such that all models have $\Gamma_{eff}=0.18$. The (blue)
squares are the data points from the Sloan Digital Sky Survey as
compiled in \cite{SDSSWMAP} and the (red) circles are
from the 2dF galaxy redshift survey as compiled in \cite{2dF}.}%
\label{entropyPk}
\end{figure}

In Fig. \ref{entropyPk} we show the baryon spectrum for
exponential and logarithmic quartessence, in the case
$\Gamma_{q}=-c_{sq}^{2}\delta_{q}/w_{q}$, for the same parameter
values of Figs. \ref{expPk} and \ref{logPk}. The results are
visually indistinguishable from the predictions of the
$\Lambda$CDM model. The same holds for the quartessence and,
therefore, the total power spectrum.

It is important to remark that the above considered entropy
perturbations are \emph{not} isocurvature ones, since there are
initial density perturbations, although no pressure fluctuations.
Moreover, if one were to consider the evolution of the coupled
baryon-photon fluid, we would set $\delta\rho
_{b}/\dot{\rho}_{b}=\delta\rho_{q}/\dot{\rho}_{q}=(3/4)\delta\rho_{r}%
/\dot{\rho}_{r}$, which is sometimes referred in the literature as
\textquotedblleft adiabatic perturbations\textquotedblright.
Therefore, the entropy perturbations needed to eliminate the
instabilities and oscillations in the power spectrum are not
straightforwardly ruled out by cosmic microwave background
radiation (CMB) observations, which have discarded purely
isocurvature modes. A complete study of the CMB spectrum in
quartessence models with entropy perturbations such as given by
Eq. (\ref{condition}) remains to be done and would be an important
test for the model.

\section{Discussion\label{disc}}

In what is currently considered the standard cosmological model,
two exotic components, DM and DE, are invoked to explain two
different phenomena: clustering of matter and cosmic acceleration.
In spite of their success in explaining most of the cosmological
observations, the exact nature of these two components remains a
mystery. In fact, so far, it has not been proven that dark matter
and dark energy are two distinct substances. This unsatisfactory
situation stimulated the search for viable quartessence
(unified-dark-matter) models \cite{scherrer,makler05}.

In this work we argue that adiabatic quartessence models with a
convex equation of state always suffer from oscillations and
suppression in the power spectrum, rendering those models
inconsistent with the data. This can be understood in very simple
terms, since, for that class of model, the maximum sound speed
always occurs at the minimum density (at which $p=-\rho$).
Therefore, the epoch of accelerated expansion is also a period of
high adiabatic sound speed. This situation is different for
equations of state that change concavity. In this case, the moment
of maximum adiabatic sound speed can be disconnected from the
start up of accelerated expansion. Such models are currently under
investigation and may provide a power spectrum in agreement with
the data.

Another alternative is to consider intrinsic entropy
perturbations. In Ref.
\cite{reis03b} it was shown that, for QCM, if $\Gamma_{q}=-c_{sq}^{2}%
\delta_{q}/w_{q}$, or equivalently, that initial conditions are
chosen such that $\delta p_{q}=0$, the oscillations and
instabilities present in the adiabatic case, disappear. In this
work, we have shown that the same type of entropy perturbations
also solves the problem of the power spectrum for both the
exponential and logarithmic quartessence models. This occurs
because, in this case the effective sound speed vanishes
($c_{s\mathrm{eff}}^{2}=\delta p/\delta\rho=0$), such that there
is no gradient term in the \textquotedblleft Euler
equation\textquotedblright\ (\ref{eq2}), and thus no oscillations
and divergences appear in the power spectrum. We therefore argue
that such entropy perturbations render generic quartessence models
in agreement with power spectrum observations. However, although
the nonadiabatic quartessence power spectrum is indistinguishable
from that of \textquotedblleft concordance\textquotedblright\
models like $\Lambda$CDM, their behavior beyond the linear regime
is quite different and, in principle, can be clearly distinguished
with weak lensing observations \cite{reis04}.

To conclude, we remark that the quartessence scenario offers an
alternative to concordance models, providing distinct predictions
that can be tested with current data. Several dark energy models,
being as \textquotedblleft exotic\textquotedblright\ as
quartessence, are nevertheless much more degenerate with the
$\Lambda$CDM model. It is therefore desirable to put some effort
into deriving observational consequences of quartessence models.

Adiabatic convex quartessence now seem to be ruled out by the
data. Nonadiabatic convex models might be discarded by weak
lensing. However, models in which the equation of state changes
its concavity are a promising alternative, even for adiabatic
perturbations \cite{makler05}. Such models deserve to be further
studied, offering potential alternatives for dark-matter and
dark-energy unification. The search for quartessence models
provides also a test for the robustness of the dark-matter plus
dark-energy paradigm. If generic unifying-dark-matter models can
be ruled out, this would give a strong support for the concordance
cosmological models.

\ack

The authors would like to thank the following Brazilian research
agencies for financial support: RRRR is partially supported by
CAPES, MM is partially supported by FAPERJ and MCT, and IW is
partially supported by CNPq. MM acknowledges the hospitality of
Fermilab, where this manuscript was completed.

\end{document}